\documentclass[twocolumn,showpacs,preprintnumbers,amsmath,amssymb,prl]{revtex4}

\usepackage{graphicx,mathrsfs,amssymb}
\usepackage{dcolumn}
\usepackage{bm}

\begin{document}

\newcommand{\trace}[1]{\ensuremath{\mathrm{Tr}\{ #1 \}}}
\newcommand{\bra}[1]{\ensuremath{\left< #1 \right|}}
\newcommand{\ket}[1]{\ensuremath{\left| #1 \right>}}
\newcommand{\ud}{\mathrm{d}}
\newcommand{\E}{\mathscr{E}}
\newcommand{\kbf}{\mathbf{k}}
\newcommand{\rbf}{\boldsymbol{r}}
\newcommand{\kappabf}{\boldsymbol{\kappa}}
\newcommand{\Ebf}{\boldsymbol{E}}
\newcommand{\mbf}[1]{\boldsymbol{#1}}
\newcommand{\aop}{\hat{a}}

\title{Tomographic reconstruction of the single-photon Fock state by
high-frequency homodyne detection}

\author{Alessandro Zavatta}

\author{Silvia Viciani}

\author{Marco Bellini}

\altaffiliation[Also at ]{
LENS and 
INFM, Florence, Italy} \email{bellini@inoa.it} \affiliation{Istituto Nazionale
di Ottica
Applicata (INOA),\\L.go E. Fermi, 6, I-50125, Florence, Italy}

\date{\today}

\begin{abstract}
A single-photon Fock state has been generated by means of conditional preparation from a
two-photon state emitted in the process of spontaneous parametric down-conversion. A recently
developed high-frequency homodyne tomography technique has been used to completely
characterize the Fock state by means of a pulse-to-pulse analysis of the detectors' difference
photocurrent. The density matrix elements of the generated state have been retrieved with a
final detection efficiency of about 57\%. A comparison has been performed between the
phase-averaged tomographic reconstructions of the Wigner function as obtained from the
measured density-matrix elements and from a direct Abel transform of the homodyne data. The
ability of our system to work at the full repetition rate of the pulsed laser (82 MHz)
substantially simplifies the detection scheme, allowing for more ``exotic'' quantum states to
be generated and analyzed.
\end{abstract}

\pacs{03.65.Wj, 03.65.Ud, 42.50.Dv}

\maketitle

\section{Introduction}

Homodyne detection is a widely used method to obtain tomographic reconstructions of classical
and quantum optical fields in the cw and pulsed regimes. It consists in mixing the unknown
signal field with a strong local oscillator (LO) field on a beam splitter whose outputs are
then detected by proportional photodetectors~\cite{fabre92}. The difference in the
photocurrents produced by the two detectors is proportional to the electric field quadrature
selected by varying the relative phase between the LO and the signal field. From a collection
of field quadratures it is then possible to reconstruct the quantum state by means of
tomographic techniques. Vacuum, thermal, coherent and squeezed states have been successfully
analyzed, in the frequency-domain, by means of a spectral analysis of the difference
photocurrent \cite{breitenbach97,vasilyev98}, and in the time domain, with the measurement of
integrated photocurrents on isolated pulses, in order to achieve a full reconstruction of the
unknown field states \cite{smithey93, crispino00,grangier04}.

Recently, a pulsed homodyne technique \cite{hansen01} has been successfully applied to the
first complete quantum state reconstruction of single-photon Fock states of the
electromagnetic field~\cite{lvovsky01}, i.e. states containing exactly one quantum of field
excitation. Such states are an important workbench for the test and the analysis of
nonclassical behavior and represent a fundamental building block for the implementation of
efficient linear quantum computation. Time-domain homodyne detection schemes as those used in
\cite{hansen01, lvovsky01} are however quite demanding from the experimental point of view,
since they require very low electronic noise and high subtraction efficiencies over very large
frequency bandwidths (from dc to a few times the repetition rate of the laser). Mostly due to
these technical limitations, all the existing applications of the latter technique have been
limited to pulse repetition rates well below the megahertz. The measurements reported in
\cite{hansen01, lvovsky01} indeed made use of a pulse picker to lower the laser repetition
frequency, and the low state production rates implied very long acquisition times in order to
store enough data for a single accurate tomographic reconstruction. More recent works by the
same group \cite{lvovsky02,lvovsky04} have adopted a different technique, where the local
oscillator pulses are sent ``on demand'' to the homodyne detector only when needed; this has
contributed to significantly increase their acquisition rates, although at the price of a more
complicated setup and lower efficiencies.

Our group has recently developed an homodyne set-up which is able to perform
high-frequency (with repetition rates up to about 100 MHz) time-domain analysis
of quantum states of light~\cite{josa02}. Its capabilities were first
demonstrated by performing a tomographic reconstruction of the Wigner function
of weak coherent states and by showing the possibility of a gated-mode
acquisition for conditional measurements. Here we present the first complete
and efficient tomographic reconstruction of a single-photon Fock state based on
such a time-domain, pulsed optical homodyne apparatus operating at the high
repetition frequencies characteristic of commonly used mode-locked laser
systems. By allowing us to perform ultrafast pulse-selective measurements, this
apparatus exhibits two fundamental advantages over alternative schemes working
at lower repetition rates: on one side, thanks to the much shorter acquisition
times, the stability requirements of the overall experimental system are
loosened; on the other, if long-term stability of the setup is available,
states exhibiting stronger non-classical features (usually characterized by
smaller generation probabilities) can be efficiently analyzed.

\section{Generation of single-photon Fock states}

A single-photon Fock state can be generated by means of a conditional state preparation on the
two-photon wavefunction emitted in the process of spontaneous parametric down-conversion
(SPDC) (see Refs.~\cite{lvovsky02:epjd,viciani04} and references therein). Such a state is
written as
\begin{eqnarray}
\ket{\Psi} &=& \ket{0}_{s}\ket{0}_{t} -
\int{\ud^3k_s\ud\omega_s \ud^3k_t \ud\omega_t }
\psi(\kbf_s, \omega_s, \kbf_t, \omega_t ) \nonumber \\
&& \times \ \ \ket{\kbf_s, \omega_s }_s
\ket{\kbf_t, \omega_t}_t,
\end{eqnarray}
where the function $\psi(\kbf_s, \omega_s, \kbf_t, \omega_t )$
describes the spatial and spectral properties of the two-photon state ({\em
biphoton}) and the subscripts $s$ and $t$ correspond to the signal and trigger
(idler) photons.

In order to non-locally select a pure state on the signal channel, trigger
photons must undergo narrow spatial and frequency filtering before being
detected by a single photon counter~\cite{ou97,lvovsky02:epjd}. The conditional non-local preparation of
the signal state is thus described by the density operator $\hat{\rho}_s$
given by the partial trace of the product between the SPDC-state density
operator and the measurement density operator $\hat{\rho}_t$:
\begin{equation}
\hat{\rho}_s = \trace{\hat{\rho}_t\ket{\Psi}\bra{\Psi}}_t
\end{equation}
where
\begin{equation}
\hat{\rho}_t = \int{\ud^3k_t\ud\omega_t T(\kbf_t,\omega_t)}
\ket{\kbf_t, \omega_t}_t \bra{\kbf_t, \omega_t }_t
\end{equation}
and $T(\kbf,\omega)$ is the transmission function of the spatial and spectral filter placed
along the trigger photon path. Note that the nonlocally-prepared signal state will only
approach a pure state if the filter transmission function $T(\kbf,\omega)$ is much narrower
than the momentum and spectral widths of the pump beam generating the SPDC pair. Accordingly,
one can define a purity parameter as $P=\mathrm{Tr}(\hat\rho_s^2)$ which, in the spectral
domain, is expressed as
\begin{equation}
P_t =
\frac{1}{\sqrt{1+\sigma_f^2/\sigma_p^2}} \label{eq:nonlocal:spectralpurity}
\end{equation}
where Gaussian profiles with widths $\sigma_p$ and $\sigma_f$ are assumed for the pump power
spectrum and for the trigger filter transmission, respectively. It is easy to see how the
spectral purity parameter $P_t$ approaches unity when $\sigma_f\ll\sigma_p$. In such a
condition the strong spectral filtering performs a non-local selection of a pure signal state
whose properties are defined by the pump as shown in \cite{bellini03,viciani04}. Similarly,
the spatial purity parameter is found to be~\cite{lvovsky02:epjd}:
\begin{equation}
P_{s} = \frac{1}{1+\kappa_i^2/\kappa_p^2}
\end{equation}
where $\kappa_i$ and $\kappa_p$ are the beam widths in the momentum space for the idler
spatial filter and for the pump, respectively.

\section{Experimental setup}

\begin{figure}[h]
\includegraphics[width=85mm]{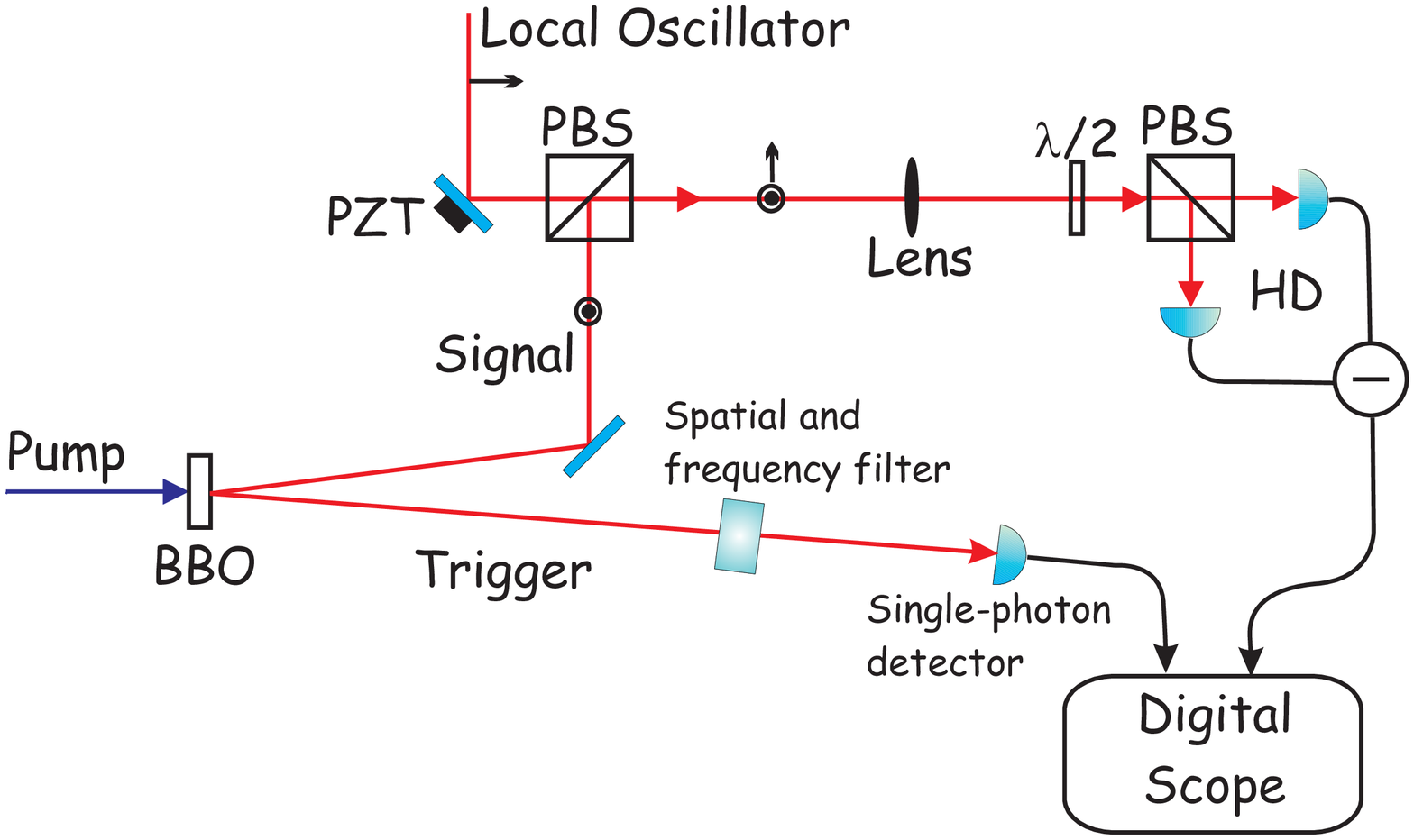}
\caption{Experimental apparatus: BBO downconverter crystal, PZT piezoelectric transducer, PBS
polarizing beamsplitter, $\lambda/2$ half-wave plate, HD homodyne detector. \label{fig:setup}}
\end{figure}

A mode-locked Ti:sapphire laser, emitting 1-2 ps long pulses at 786 nm with a repetition rate
of 82 MHz is used as the primary source for the experiment, schematically drawn in
Fig.\ref{fig:setup}. The laser pulses are first frequency doubled in a 13-mm long LBO crystal
which produces the pump pulses centered at 393 nm, with a mean power of about 100 mW. The UV
pump pulses, after a spatial mode cleaner, are then slightly focused, with a beam waist $w_p =
220\ \mu\textrm{m}$, and the waist position is carefully located inside a 3-mm thick, Type~I
BBO crystal cut for collinear degenerate SPDC. The crystal is adjusted in order to obtain a
signal-trigger cone beam with an angle of $\sim 3^\circ$. Signal and trigger photon pairs are
then selected by means of irises placed at 70 cm from the output face of the crystal.

The trigger beam is directed to the state preparation channel, where it is spectrally filtered
by means of a pair of etalon interference filters and is then coupled into a single-mode fiber
connected to a single photon counting module with an active-area diameter of $200\
\mu\textrm{m}$ (Perkin-Elmer SPCM AQR-14). The typical trigger count rate after spectral and
spatial filtering is of about 500 s$^{-1}$, with a dark count rate of the order of 10 s$^{-1}$
when both are gated with the laser pulses.

The combination of the two etalon cavities yields a final spectral width for the trigger
channel of about 50 GHz. By comparing this with a measured pump width of 430 GHz, we estimate
a spectral purity parameter of $P_t=0.98$. However, due to the second harmonic generation
process, the LO spectral width remains $\sqrt{2}$ times narrower than that of the selected
single-photon state. The degree of mode-matching between the LO and the single-photon state
cannot be improved by filtering the LO beam and might be increased only by inserting a
spectral filter on the UV pump. However, this solution would also drastically lower the count
rate on the trigger detector and has not been implemented yet.

Filtering in the spatial domain is achieved by means of the single-mode fiber and our
experimental parameters lead to a spatial purity parameter $P_s=0.86$, which again means that
the conditionally prepared single-photon state mode is essentially determined by the pump
characteristics. The LO beam is spatially mode-matched to the fiber-selected signal photon
mode by the insertion of appropriate lens combinations along its path. In order to finely
adjust the alignment and the synchronization between the signal and LO pulses, we use the
stimulated beam produced by injecting a seed pulse into the crystal. Under appropriate
conditions~\cite{lvovsky02:epjd}, the amplified seeded beam can spatially well simulate the
selected signal beam and can be used for alignment purposes.

The LO pulses are obtained by splitting a small portion of the laser emission, and their
polarization is rotated by a half-wave plate in order to overlap them with the
conditionally-prepared single-photon signal beam onto a polarizing beam splitter (PBS). One of
the steering mirrors is mounted on a piezoelectric transducer in order to vary and control the
LO phase. A 300-mm focal-length lens, an additional $\lambda/2$ wave plate, and a second
polarizing beam splitter are then used to accurately mix the two field modes and focus them
onto the two photodiodes of the homodyne detector.

\section{High-frequency balanced homodyne detection}
Two photodiodes (Hamamatsu S3883, with active area 1.7 mm$^2$) are connected to the positive
and negative inputs of an operational amplifier providing a gain of about 10. The bias voltage
of the two photodiodes is optimized in order to provide the maximum extinction in the
difference signal when the optical power impinging on the two is well balanced. The amplifier
output is AC-coupled (cut-off frequency of 1 MHz at -3 dB) to a second amplifier with gain G
=100 and bandwidth extending from dc to 300 MHz, before being sent to a digital scope with an
analog bandwidth of 1 GHz, a sampling rate of 10 Gsamples/s and a vertical resolution of 8
bits.

In order to decrease the effect of the dark counts in the single-photon trigger detector, a
strict coincidence with the signal coming from the laser pulse train is used as the trigger
for the acquisition of the homodyne signal. Although this slightly reduces the trigger count
rate, it is effective in increasing the ratio of ``true'' to ``false'' trigger events to more
than 99\%.

A single temporal frame composed of 250 acquisition points is stored in the oscilloscope
memory whenever a trigger event occurs. Each of these frames contains two consecutive LO
pulses where only the first one is coincident with the detection of a trigger photon and
contains the ``single-photon information'', while the second one can be used for the
measurement of the reference vacuum state. About 5000 frames can be stored in a sequence at a
maximum rate of 160,000 frames per second. With an average rate of trigger counts of about 300
s$^{-1}$, this means that each sequence is acquired in about 15-20 s. Each sequence of frames
is then transferred to a personal computer where the areas of the pulses are measured and
their statistic distributions are analyzed in real time. The acquisition of two nearby LO
pulses in each frame allows the simultaneous analysis of the ``signal'' pulses, namely those
that have interfered with the signal photons, and of the ``dark'' pulses, so that both the
Fock-state and the LO shot-noise distributions are immediately available from a single
sequence, independent from low-frequency fluctuations in the system. The acquisition and
analysis of a sequence of 5000 signal and dark pulses yields an experimental Wigner marginal
distribution in just about 30 s.

It is interesting to compare these acquisition times with those reported in \cite{lvovsky01}
by the Constance group for the first tomographic reconstruction of the single-photon Fock
state. Due to the limited bandwidth of their homodyne detector (1 MHz), the overall pulse rate
had to be lowered to about 800 kHz by means of a pulse picker at the exit of the laser. In
that case about 12,000 experimental data points were acquired in a 14-hour experimental run in
order to obtain an accurate phase-averaged marginal distribution of the quantum state. Our
setup now allows the same kind of acquisition to be performed in just about one minute, with a
gain of almost three orders of magnitude in the measurement time. It is clear that such a
speed-up can help to loosen some of the constraints on the overall stability of the generation
system and, in particular, on the interferometric stability of the beam paths which is needed
when phase-dependent quantum states are to be analyzed. On the other hand, by greatly
increasing the overall number of available LO pulses in a reasonable experimental time
interval, our scheme will allow us to investigate much rarer and more exotic events, such as
those involving higher photon numbers \cite{josa02,montina98}.

Note that recent experiments from the same group \cite{lvovsky02,lvovsky04} have seen the
adoption of a clever solution to the limited acquisition rate caused by the low detector
bandwidth. Instead of reducing the whole experimental repetition rate, the pulse picker has
been placed only in the LO path and activated ``on demand'' upon a trigger detection event.
This has allowed to keep reasonable pair production rates and to obtain a final acquisition
rate similar to the one presented here. However, the introduction of the pulse picker, besides
significantly complicating the experimental setup, has the important drawback that the trigger
delay of the shutter in the LO path has to be compensated by a long (about 15 m) optical delay
line on the signal photon path in order for the two to reach the mixing beam-splitter at the
same time. This clearly introduces additional losses and may degrade both the mode quality and
the accurate phase stability of the fragile quantum state.

The performances of our detector were verified with a preliminary measurement of the laser
shot-noise by analyzing the variances of the ``dark'' pulse areas as a function of the LO
power. A linear behavior is preserved over a wide range of LO powers, extending up to 9 mW,
and with a signal to electronic-noise ratio of about 12 dB when the device is operated at the
optimum LO power of 7 mW.

\section{Quantum state reconstruction}
Balanced homodyne detection allows the measurement of the signal electric field quadratures
$x_\theta$ as a function of the relative phase $\theta$ imposed between the LO and the signal.
By performing a series of homodyne measurements on equally-prepared states it is possible to
obtain the probability distributions $p(x,\theta)$ of the quadrature operator $\hat x_\theta$
that correspond to the marginals of the Wigner quasi-probability distribution
$W(x,y)$~\cite{vogel89}:
\begin{equation}
p(x,\theta) =\int_{-\infty}^{+\infty} W(x\cos \theta-y\sin\theta,
x\sin \theta+y\cos\theta)\ud y. \label{eq:radon}
\end{equation}

In principle, given a sufficient number of quadrature distributions for different values of
$\theta$, one is able to reconstruct the Wigner function by inverting Eq.~(\ref{eq:radon}). In
the case of Fock states, some simplification in the reconstruction procedures can be
accomplished thanks to the phase-invariant nature of the Wigner function~\cite{lvovsky01}.
Only one phase-averaged marginal distribution is required for the reconstruction in this case
and, from the experimental point of view, this also means that the stabilization of the LO
phase is not needed. Furthermore, the Abel transform, relying on the cylindrical symmetry of
the problem, can be used instead of the Radon transform in order to reconstruct a section of
the Wigner function $W(r)$~\cite{leonhardt97}, as given by
\begin{equation}
W(r) = -\frac{1}{\pi} \int_r^{+\infty} \frac{\partial p_{\textrm{av}}(x)}{\partial x}
\frac{\ud x}{\sqrt{x^2-r^2}}
\end{equation}
where $p_{\textrm{av}}(x)$ is the phase-averaged marginal distribution. The numerical
implementation of this method is however extremely sensitive to noise and a large number of
data is required to obtain a reliable result. Since this reconstruction procedure involves the
application of numerically-unreliable methods, the group of D'Ariano has developed an
alternative precise technique that permits the reconstruction of the density matrix elements
directly from the homodyne data~\cite{dariano94}.

\begin{figure}
\includegraphics[width=80mm]{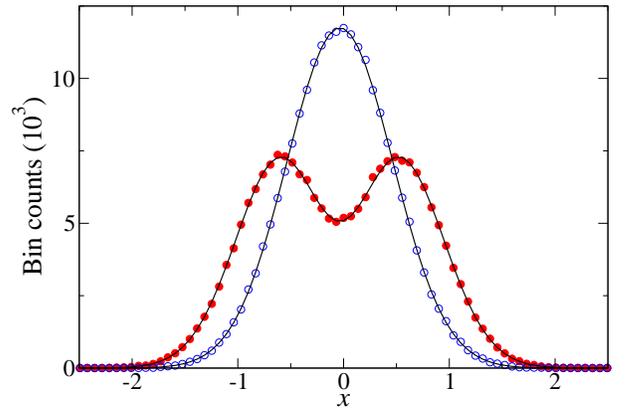}
\caption{Vacuum (open circles) and signal (filled circles) phase-averaged quadrature field
distributions. The solid lines are fitted curves. The quadrature is normalized to the vacuum.
\label{fig:marginal}}
\end{figure}

The distributions of the acquired pulse areas for the vacuum and the single-photon state are
shown in Fig.~\ref{fig:marginal},
where each histogram is obtained from the analysis of about 200,000 acquired pulses. Such
distributions represent the marginal integrals of the corresponding field state. The non-unity
detection efficiency of the apparatus prevents us from observing the real single-photon Wigner
function, and what we get instead is its convolution with the vacuum one. The convolution
result is the well known $s$-parametrized quasi-probability distribution with the $s$
parameter scaled by the detection efficiency $\eta$~\cite{mandelwolf,leonhardt97}. The
expression for the corresponding marginal distribution is:
\begin{equation}
p(x;\eta) = \sqrt{\frac{2}{\pi}} [1-\eta(1-4x^2)] e^{-2x^2}
\end{equation}
From a fit of the experimental distributions to the corresponding theoretical curves we obtain
an overall detection efficiency of $\eta = 0.574 \pm 0.002$.

The expected overall efficiency $\eta$ is given by the contribution of several terms:
\begin{equation}
\eta = \eta_{\textrm{hd}}\eta_{\textrm{dc}}\eta_{\textrm{mm}}
\end{equation}
where $\eta_{\textrm{hd}} = 0.90$ is the efficiency of the homodyne detector including the PBS
losses, while $\eta_{\textrm{dc}}=0.99$ is the efficiency connected to the dark counts of the
trigger detector. The mode-matching efficiency is given by
$\eta_{\textrm{mm}}=\eta_{\textrm{exp}}\sqrt{P_sP_t}$, where $\sqrt{P_sP_t}$ depends on the
tightness of the spatial and spectral filtering on the trigger channel and constitutes the
upper limit for the mode matching efficiency achievable between a pure state (LO) and the
non-locally selected single photon state. The term $\eta_{\textrm{exp}}$ is connected with the
non-ideal experimental conditions and can be estimated from the visibility of the interference
fringes formed between the LO and the stimulated radiation used for
alignment~\cite{lvovsky02:epjd}. We obtain $\eta_{\textrm{exp}} \simeq 0.7$ and the overall
efficiency is thus expected to be $\eta \simeq 0.6$, in good agreement with the value
retrieved from the fit of the marginal data to the theoretical distributions.

\begin{figure}
\begin{center}
\includegraphics*[width=80mm]{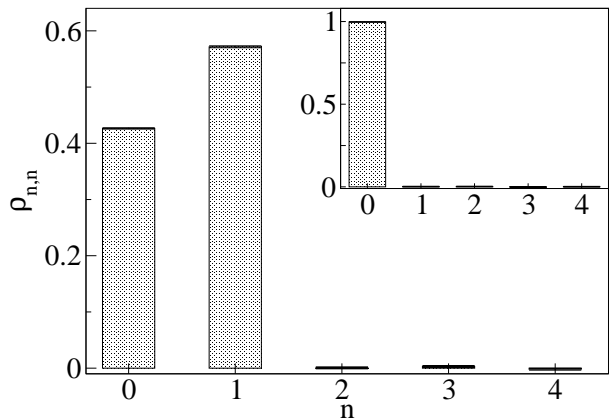}
\end{center}
\caption{Reconstructed diagonal elements of the density matrix for the single-photon state and
the vacuum (inset). \label{fig:dm}}
\end{figure}

In Fig.~\ref{fig:dm}
the reconstructed diagonal elements $\rho_{n,n}$ of the density matrix are reported as
obtained by averaging the so-called pattern functions over our data:
\begin{equation}
\rho_{n,n} = \bra{n}\hat\rho\ket{n} = \pi \int_{-\infty}^{+\infty} p_{\textrm{av}}(x)
f_{nn}(x) \ud x.
\end{equation}
Here $f_{nn}(x)$ are the factorized pattern functions that can be numerically implemented
following the numerical recipes given in Ref.~\cite{dariano97,leonhardt97}.
Figure~\ref{fig:dm} clearly shows the non-poissonian photon distribution of the reconstructed
state, which is a statistical mixture of the single-photon Fock state with the vacuum state
$\hat \rho = \rho_{0,0}\ket{0}\bra{0} + \rho_{1,1} \ket{1}\bra{1}$ with $\rho_{0,0}=0.426 \pm
0.001$ and $\rho_{1,1}=0.572 \pm 0.002$ in good agreement with the efficiency value found by
fitting the marginal distributions.

\begin{figure}
\begin{center}
\includegraphics*[width=85mm]{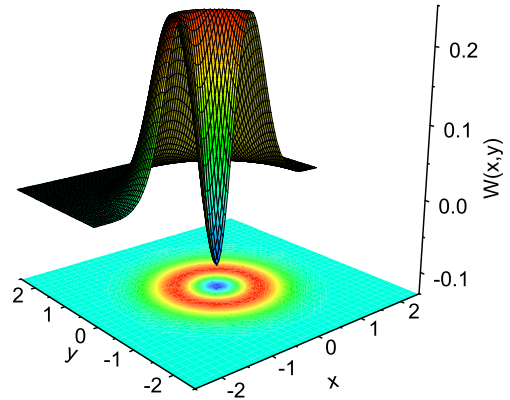}
\end{center}
\caption{Wigner function of the single-photon Fock state as obtained from the reconstructed
density-matrix elements. The negativity of the distribution, a clear proof of the
non-classical character of the state, is evident around the origin. \label{fig:wigner}}
\end{figure}

As discussed by D'Ariano in \cite{dariano97,dariano95}, the Wigner function can be obtained from the
reconstructed density-matrix elements in a dimensionally-truncated Hilbert space as:
\begin{equation}
W(x,y)=\textrm{Re}\sum_{d=0}^{M}e^{id\arctan(y/x)}\sum_{n=0}^{M-d}\Lambda(n,d,x,y)\rho_{n,n+d}
\label{eq:wigdar}
\end{equation}
and
\begin{eqnarray}
\Lambda(n,d,x,y)=\frac{2(-1)^n}{\pi}[2-\delta_{d0}]|2(x+iy)|^d \nonumber \\
\times\sqrt{\frac{n!}{(n+d)!}}e^{-2(x^2+y^2)}L_n^d[4(x^2+y^2)]
\end{eqnarray}
where $L_n^d(x)$ are Laguerre polynomials, $\delta$ is the Kronecker delta and $M$ is the
maximal quantum number of the reconstructed density matrix. In Fig.\ref{fig:wigner}
we show the Wigner function obtained from the first 10 reconstructed diagonal
elements of the density matrix and by setting all the off-diagonal terms to
zero thanks to the phase-invariance of the state. The reconstructed Wigner
function assumes negative values at the origin, showing the non-classical
features of the single-photon state. We find $W_{\textrm{meas}}(0,0)=-0.0995$,
which is quite close to the theoretical value of the Wigner function
$W(0,0;\eta)=-0.0942$ obtained by using Eq.~(\ref{eq:wigdar}) with the mixed state
defined by
\begin{equation}
\hat \rho = (1-\eta)\ket{0}\bra{0} + \eta \ket{1}\bra{1} \label{mix}
\end{equation}
with the efficiency value ($\eta=0.574$) retrieved from the fit of the phase-averaged marginal
distributions.

\begin{figure}
\includegraphics[width=75mm]{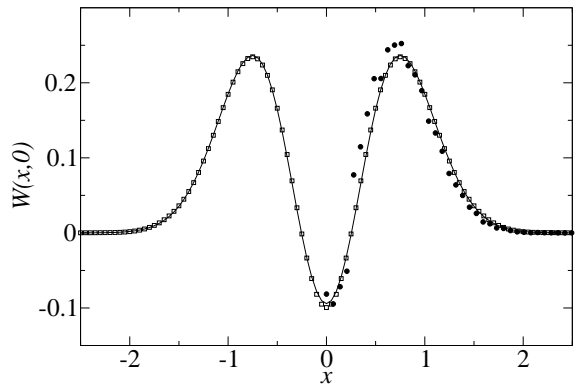}
\caption{$W(x,0)$ section of the reconstructed Wigner function. Filled circles: Abel
transform; empty squares: section of Fig.\ref{fig:wigner}; solid line: section corresponding
to the mixed state of Eq.~(\ref{mix}).\label{fig:rec_methods}}
\end{figure}

A comparison between the different methods of Wigner function reconstruction is presented in
Fig.~\ref{fig:rec_methods} where the $W(x,0)$ section is plotted. Both the profile obtained by
a direct implementation of the Abel transform on the acquired homodyne histograms using the
basic algorithm of Nestor and Olsen~\cite{cremers66} and a section of the Wigner function as
reported in Fig.\ref{fig:wigner} are shown, together with the curve retrieved using the mixed
state of Eq.~(\ref{mix}). It is evident that the direct reconstruction method based on the Abel
transform is extremely sensitive to the noise present in the data and fails to accurately
reproduce the expected vacuum-convoluted single-photon Wigner function. On the other hand, the
reconstruction via the density-matrix elements is much smoother and extremely close to the
expected profile.

Recently, alternative reconstruction techniques based on maximum-likelihood estimation methods
have been proposed~\cite{banaszek99,lvovsky03} and used to analyze experimental
data~\cite{lvovsky04}. Such methods can reconstruct the density matrix elements and the Wigner
function with lower requirements on the number of data samples without a substantial increase
in the statistical errors. The implementation of these methods might contribute to further
decrease the acquisition time of our experiment whenever this becomes a limiting factor, i.e.
when more exotic quantum states are analyzed.

\section{Conclusions}

We have applied a recently developed technique of time-domain, high-frequency homodyne
detection to perform quantum tomographic reconstructions of the single-photon Fock state. The
scheme for the generation of the quantum field is based on remote state preparation from an
entangled two-photon wavefunction emitted in the process of pulsed parametric down-conversion.
By analyzing pulsed homodyne signals at the full repetition rate of the mode-locked source
laser, this scheme allows us to perform full tomographic reconstructions in relatively short
times. Furthermore, by allowing us to avoid the introduction of additional mechanisms of pulse
selection, our scheme has the potential to achieve more compact realizations (which may be an
issue when an accurate control of the relative phase in needed) and higher detection
efficiencies in the analysis of quantum states. Finally, we have shown that the methods of
Wigner function retrieval based on the direct analysis of the homodyne data are too sensitive
to experimental noise and that much more accurate results can be obtained from the
reconstructed density-matrix elements.

\section{Acknowledgments}
This work has been partially supported by the Italian Ministry of University and Scientific
Research (MIUR), under the FIRB contract RBNE01KZ94. The authors would also like to thank the
Physics Department of the University of Florence for the kind hospitality and P.~Poggi of the
electronic workshop at INOA for the skillful implementation of the high-frequency detection
electronics.

\bibliography{Fock_bib}

\begin{thebibliography}{25}
\expandafter\ifx\csname natexlab\endcsname\relax\def\natexlab#1{#1}\fi
\expandafter\ifx\csname bibnamefont\endcsname\relax
  \def\bibnamefont#1{#1}\fi
\expandafter\ifx\csname bibfnamefont\endcsname\relax
  \def\bibfnamefont#1{#1}\fi
\expandafter\ifx\csname citenamefont\endcsname\relax
  \def\citenamefont#1{#1}\fi
\expandafter\ifx\csname url\endcsname\relax
  \def\url#1{\texttt{#1}}\fi
\expandafter\ifx\csname urlprefix\endcsname\relax\def\urlprefix{URL }\fi
\providecommand{\bibinfo}[2]{#2}
\providecommand{\eprint}[2][]{\url{#2}}

\bibitem[{\citenamefont{Reynaud et~al.}(1992)\citenamefont{Reynaud, Heidmann,
  Giacobono, and Fabre}}]{fabre92}
\bibinfo{author}{\bibfnamefont{S.}~\bibnamefont{Reynaud}},
  \bibinfo{author}{\bibfnamefont{A.}~\bibnamefont{Heidmann}},
  \bibinfo{author}{\bibfnamefont{E.}~\bibnamefont{Giacobono}},
  \bibnamefont{and} \bibinfo{author}{\bibfnamefont{C.}~\bibnamefont{Fabre}}, in
  \emph{\bibinfo{booktitle}{Progress in Optics}}, edited by
  \bibinfo{editor}{\bibfnamefont{E.}~\bibnamefont{Wolf}}
  (\bibinfo{publisher}{Elsevier}, \bibinfo{address}{Amsterdam},
  \bibinfo{year}{1992}), vol.~\bibinfo{volume}{30}, p.~\bibinfo{pages}{1}.

\bibitem[{\citenamefont{Breitenbach et~al.}(1997)\citenamefont{Breitenbach,
  Schiller, and Mlynek}}]{breitenbach97}
\bibinfo{author}{\bibfnamefont{G.}~\bibnamefont{Breitenbach}},
  \bibinfo{author}{\bibfnamefont{S.}~\bibnamefont{Schiller}}, \bibnamefont{and}
  \bibinfo{author}{\bibfnamefont{J.}~\bibnamefont{Mlynek}},
  \bibinfo{journal}{Nature} \textbf{\bibinfo{volume}{387}},
  \bibinfo{pages}{471} (\bibinfo{year}{1997}).

\bibitem[{\citenamefont{Vasilyev et~al.}(1998)\citenamefont{Vasilyev, Choi,
  Kumar, and D'Ariano}}]{vasilyev98}
\bibinfo{author}{\bibfnamefont{M.}~\bibnamefont{Vasilyev}},
  \bibinfo{author}{\bibfnamefont{S.-K.} \bibnamefont{Choi}},
  \bibinfo{author}{\bibfnamefont{P.}~\bibnamefont{Kumar}}, \bibnamefont{and}
  \bibinfo{author}{\bibfnamefont{G.~M.} \bibnamefont{D'Ariano}},
  \bibinfo{journal}{Optics\ Lett.} \textbf{\bibinfo{volume}{23}},
  \bibinfo{pages}{1393} (\bibinfo{year}{1998}).

\bibitem[{\citenamefont{Smithey et~al.}(1993)\citenamefont{Smithey, Beck,
  Raymer, and Faridani}}]{smithey93}
\bibinfo{author}{\bibfnamefont{D.~T.} \bibnamefont{Smithey}},
  \bibinfo{author}{\bibfnamefont{M.}~\bibnamefont{Beck}},
  \bibinfo{author}{\bibfnamefont{M.~G.} \bibnamefont{Raymer}},
  \bibnamefont{and} \bibinfo{author}{\bibfnamefont{A.}~\bibnamefont{Faridani}},
  \bibinfo{journal}{Phys.\ Rev.\ Lett.} \textbf{\bibinfo{volume}{70}},
  \bibinfo{pages}{1244} (\bibinfo{year}{1993}).

\bibitem[{\citenamefont{Crispino et~al.}(2000)\citenamefont{Crispino, Giuseppe,
  \mbox{De Martini}, and Mataloni}}]{crispino00}
\bibinfo{author}{\bibfnamefont{M.}~\bibnamefont{Crispino}},
  \bibinfo{author}{\bibfnamefont{G.~D.} \bibnamefont{Giuseppe}},
  \bibinfo{author}{\bibfnamefont{F.}~\bibnamefont{\mbox{De Martini}}},
  \bibnamefont{and} \bibinfo{author}{\bibfnamefont{P.}~\bibnamefont{Mataloni}},
  \bibinfo{journal}{Fortschr.\ Phys.} \textbf{\bibinfo{volume}{48}},
  \bibinfo{pages}{589} (\bibinfo{year}{2000}).

\bibitem[{\citenamefont{Wenger et~al.}(2004)\citenamefont{Wenger,
  Tualle-Brouri, and Grangier}}]{grangier04}
\bibinfo{author}{\bibfnamefont{J.}~\bibnamefont{Wenger}},
  \bibinfo{author}{\bibfnamefont{R.}~\bibnamefont{Tualle-Brouri}},
  \bibnamefont{and} \bibinfo{author}{\bibfnamefont{P.}~\bibnamefont{Grangier}},
  \bibinfo{journal}{Phys.\ Rev.\ Lett.} \textbf{\bibinfo{volume}{92}},
  \bibinfo{pages}{153601} (\bibinfo{year}{2004}).

\bibitem[{\citenamefont{Hansen et~al.}(2001)\citenamefont{Hansen, Aichele,
  Hettich, Lodahl, Lvovsky, Mlynek, and Schiller}}]{hansen01}
\bibinfo{author}{\bibfnamefont{H.}~\bibnamefont{Hansen}},
  \bibinfo{author}{\bibfnamefont{T.}~\bibnamefont{Aichele}},
  \bibinfo{author}{\bibfnamefont{C.}~\bibnamefont{Hettich}},
  \bibinfo{author}{\bibfnamefont{P.}~\bibnamefont{Lodahl}},
  \bibinfo{author}{\bibfnamefont{A.~I.} \bibnamefont{Lvovsky}},
  \bibinfo{author}{\bibfnamefont{J.}~\bibnamefont{Mlynek}}, \bibnamefont{and}
  \bibinfo{author}{\bibfnamefont{S.}~\bibnamefont{Schiller}},
  \bibinfo{journal}{Opt.\ Lett.} \textbf{\bibinfo{volume}{26}},
  \bibinfo{pages}{1714} (\bibinfo{year}{2001}).

\bibitem[{\citenamefont{Lvovsky et~al.}(2001)\citenamefont{Lvovsky, Hansen,
  Aichele, Benson, Mlynek, and Schiller}}]{lvovsky01}
\bibinfo{author}{\bibfnamefont{A.~I.} \bibnamefont{Lvovsky}},
  \bibinfo{author}{\bibfnamefont{H.}~\bibnamefont{Hansen}},
  \bibinfo{author}{\bibfnamefont{T.}~\bibnamefont{Aichele}},
  \bibinfo{author}{\bibfnamefont{O.}~\bibnamefont{Benson}},
  \bibinfo{author}{\bibfnamefont{J.}~\bibnamefont{Mlynek}}, \bibnamefont{and}
  \bibinfo{author}{\bibfnamefont{S.}~\bibnamefont{Schiller}},
  \bibinfo{journal}{Phys.\ Rev.\ Lett.} \textbf{\bibinfo{volume}{87}},
  \bibinfo{pages}{050402} (\bibinfo{year}{2001}).

\bibitem[{\citenamefont{Lvovsky and Shapiro}(2002)}]{lvovsky02}
\bibinfo{author}{\bibfnamefont{A.~I.} \bibnamefont{Lvovsky}} \bibnamefont{and}
  \bibinfo{author}{\bibfnamefont{J.~H.} \bibnamefont{Shapiro}},
  \bibinfo{journal}{Phys.\ Rev.\ A} \textbf{\bibinfo{volume}{65}},
  \bibinfo{pages}{033830} (\bibinfo{year}{2002}).

\bibitem[{\citenamefont{Babichev et~al.}(2004)\citenamefont{Babichev, Brezger,
  and Lvovsky}}]{lvovsky04}
\bibinfo{author}{\bibfnamefont{S.~A.} \bibnamefont{Babichev}},
  \bibinfo{author}{\bibfnamefont{B.}~\bibnamefont{Brezger}}, \bibnamefont{and}
  \bibinfo{author}{\bibfnamefont{A.~I.} \bibnamefont{Lvovsky}},
  \bibinfo{journal}{Phys.\ Rev.\ Lett.} \textbf{\bibinfo{volume}{92}},
  \bibinfo{pages}{047903} (\bibinfo{year}{2004}).

\bibitem[{\citenamefont{Zavatta et~al.}(2002)\citenamefont{Zavatta, Bellini,
  Ramazza, Marin, and Arecchi}}]{josa02}
\bibinfo{author}{\bibfnamefont{A.}~\bibnamefont{Zavatta}},
  \bibinfo{author}{\bibfnamefont{M.}~\bibnamefont{Bellini}},
  \bibinfo{author}{\bibfnamefont{P.~L.} \bibnamefont{Ramazza}},
  \bibinfo{author}{\bibfnamefont{F.}~\bibnamefont{Marin}}, \bibnamefont{and}
  \bibinfo{author}{\bibfnamefont{F.~T.} \bibnamefont{Arecchi}},
  \bibinfo{journal}{J.\ Opt.\ Soc.\ Am.\ B} \textbf{\bibinfo{volume}{19}},
  \bibinfo{pages}{1189} (\bibinfo{year}{2002}).

\bibitem[{\citenamefont{Aichele et~al.}(2002)\citenamefont{Aichele, Lvovsky,
  and Schiller}}]{lvovsky02:epjd}
\bibinfo{author}{\bibfnamefont{T.}~\bibnamefont{Aichele}},
  \bibinfo{author}{\bibfnamefont{A.~I.} \bibnamefont{Lvovsky}},
  \bibnamefont{and} \bibinfo{author}{\bibfnamefont{S.}~\bibnamefont{Schiller}},
  \bibinfo{journal}{Eur.\ Phys.\ J.\ D} \textbf{\bibinfo{volume}{18}},
  \bibinfo{pages}{237} (\bibinfo{year}{2002}).

\bibitem[{\citenamefont{Viciani et~al.}(2004)\citenamefont{Viciani, Zavatta,
  and Bellini}}]{viciani04}
\bibinfo{author}{\bibfnamefont{S.}~\bibnamefont{Viciani}},
  \bibinfo{author}{\bibfnamefont{A.}~\bibnamefont{Zavatta}}, \bibnamefont{and}
  \bibinfo{author}{\bibfnamefont{M.}~\bibnamefont{Bellini}},
  \bibinfo{journal}{Phys.\ Rev.\ A} \textbf{\bibinfo{volume}{69}},
  \bibinfo{pages}{053801} (\bibinfo{year}{2004}).

\bibitem[{\citenamefont{Ou}(1997)}]{ou97}
\bibinfo{author}{\bibfnamefont{Z.~Y.} \bibnamefont{Ou}},
  \bibinfo{journal}{Quantum\ Semiclass.\ Opt.} \textbf{\bibinfo{volume}{9}},
  \bibinfo{pages}{599} (\bibinfo{year}{1997}).

\bibitem[{\citenamefont{Bellini et~al.}(2003)\citenamefont{Bellini, Marin,
  Viciani, Zavatta, and Arecchi}}]{bellini03}
\bibinfo{author}{\bibfnamefont{M.}~\bibnamefont{Bellini}},
  \bibinfo{author}{\bibfnamefont{F.}~\bibnamefont{Marin}},
  \bibinfo{author}{\bibfnamefont{S.}~\bibnamefont{Viciani}},
  \bibinfo{author}{\bibfnamefont{A.}~\bibnamefont{Zavatta}}, \bibnamefont{and}
  \bibinfo{author}{\bibfnamefont{F.~T.} \bibnamefont{Arecchi}},
  \bibinfo{journal}{Phys.\ Rev.\ Lett.} \textbf{\bibinfo{volume}{90}},
  \bibinfo{pages}{043602} (\bibinfo{year}{2003}).

\bibitem[{\citenamefont{Montina and Arecchi}(1998)}]{montina98}
\bibinfo{author}{\bibfnamefont{A.}~\bibnamefont{Montina}} \bibnamefont{and}
  \bibinfo{author}{\bibfnamefont{F.~T.} \bibnamefont{Arecchi}},
  \bibinfo{journal}{Phys.\ Rev. A} \textbf{\bibinfo{volume}{58}},
  \bibinfo{pages}{3472} (\bibinfo{year}{1998}).

\bibitem[{\citenamefont{Vogel and Risken}(1989)}]{vogel89}
\bibinfo{author}{\bibfnamefont{K.}~\bibnamefont{Vogel}} \bibnamefont{and}
  \bibinfo{author}{\bibfnamefont{H.}~\bibnamefont{Risken}},
  \bibinfo{journal}{Phys.\ Rev. A} \textbf{\bibinfo{volume}{40}},
  \bibinfo{pages}{2847} (\bibinfo{year}{1989}).

\bibitem[{\citenamefont{Leonhardt}(1997)}]{leonhardt97}
\bibinfo{author}{\bibfnamefont{U.}~\bibnamefont{Leonhardt}},
  \emph{\bibinfo{title}{Measuring the quantum state of light}}
  (\bibinfo{publisher}{Cambridge University Press},
  \bibinfo{address}{Cambridge, England}, \bibinfo{year}{1997}).

\bibitem[{\citenamefont{D'Ariano et~al.}(1994)\citenamefont{D'Ariano,
  Macchiavello, and Paris}}]{dariano94}
\bibinfo{author}{\bibfnamefont{G.~M.} \bibnamefont{D'Ariano}},
  \bibinfo{author}{\bibfnamefont{C.}~\bibnamefont{Macchiavello}},
  \bibnamefont{and} \bibinfo{author}{\bibfnamefont{M.~G.~A.}
  \bibnamefont{Paris}}, \bibinfo{journal}{Phys.\ Rev.\ A}
  \textbf{\bibinfo{volume}{50}}, \bibinfo{pages}{4298} (\bibinfo{year}{1994}).

\bibitem[{\citenamefont{Mandel and Wolf}(1995)}]{mandelwolf}
\bibinfo{author}{\bibfnamefont{L.}~\bibnamefont{Mandel}} \bibnamefont{and}
  \bibinfo{author}{\bibfnamefont{E.}~\bibnamefont{Wolf}},
  \emph{\bibinfo{title}{Optical Coherence and quantum optics}}
  (\bibinfo{publisher}{Cambridge University Press},
  \bibinfo{address}{Cambridge, England}, \bibinfo{year}{1995}).

\bibitem[{\citenamefont{D'Ariano}(1997)}]{dariano97}
\bibinfo{author}{\bibfnamefont{G.~M.} \bibnamefont{D'Ariano}}, in
  \emph{\bibinfo{booktitle}{Quantum Optics and Spectroscopy of Solids}}, edited
  by \bibinfo{editor}{\bibfnamefont{T.}~\bibnamefont{Hakio\v{g}lu}}
  \bibnamefont{and} \bibinfo{editor}{\bibfnamefont{A.}~\bibnamefont{Shumovsky}}
  (\bibinfo{publisher}{Kluwer Academic Publishers}, \bibinfo{year}{1997}), pp.
  \bibinfo{pages}{175--202}.

\bibitem[{\citenamefont{D'Ariano}(1995)}]{dariano95}
\bibinfo{author}{\bibfnamefont{G.~M.} \bibnamefont{D'Ariano}},
  \bibinfo{journal}{Quantum.\ Semiclass.\ Opt.} \textbf{\bibinfo{volume}{7}},
  \bibinfo{pages}{693} (\bibinfo{year}{1995}).

\bibitem[{\citenamefont{Cremers and Birkebak}(1966)}]{cremers66}
\bibinfo{author}{\bibfnamefont{C.~J.} \bibnamefont{Cremers}} \bibnamefont{and}
  \bibinfo{author}{\bibfnamefont{C.}~\bibnamefont{Birkebak}},
  \bibinfo{journal}{Appl.\ Opt.} \textbf{\bibinfo{volume}{5}},
  \bibinfo{pages}{1057} (\bibinfo{year}{1966}).

\bibitem[{\citenamefont{Banaszek et~al.}(1999)\citenamefont{Banaszek, D'Ariano,
  Paris, and Sacchi}}]{banaszek99}
\bibinfo{author}{\bibfnamefont{K.}~\bibnamefont{Banaszek}},
  \bibinfo{author}{\bibfnamefont{G.~M.} \bibnamefont{D'Ariano}},
  \bibinfo{author}{\bibfnamefont{M.~G.~A.} \bibnamefont{Paris}},
  \bibnamefont{and} \bibinfo{author}{\bibfnamefont{M.~F.}
  \bibnamefont{Sacchi}}, \bibinfo{journal}{Phys.\ Rev.\ A}
  \textbf{\bibinfo{volume}{61}}, \bibinfo{pages}{010304}
  (\bibinfo{year}{1999}).

\bibitem[{\citenamefont{Lvovsky}(2003)}]{lvovsky03}
\bibinfo{author}{\bibfnamefont{A.~I.} \bibnamefont{Lvovsky}}
  (\bibinfo{year}{2003}), \eprint{quant-ph/0311097}.

\end{thebibliography}

\end{document}